\documentclass[12pt]{article}

\usepackage{cite}
\usepackage{epsfig}
\usepackage{amsmath,amssymb}
\usepackage{array}
\usepackage{multirow}
\usepackage{pstricks}

\usepackage{color}

\def\mathswitch#1{\relax\ifmmode#1\else$#1$\fi}
\def\mathswitchr#1{\relax\ifmmode{\mathrm{#1}}\else$\mathrm{#1}$\fi}
\newcommand{\PW}{\mathswitchr W}
\newcommand{\PZ}{\mathswitchr Z}

\newcommand{\Pb}{\mathswitchr b}
\newcommand{\Pt}{\mathswitchr t}

\newcommand{\MZ}{\mathswitch {m_\PZ}}

\newcommand{\gev}{\,\, \mathrm{GeV}}

\newcommand{\lesim}{\,\raisebox{-.1ex}{$_{\textstyle<}\atop^{\textstyle\sim}$}\,}
\newcommand{\gesim}{\,{_{\textstyle
>}\atop^{\textstyle\sim}}\,}

\newcommand{\mycaption}[1]{\caption{\sl #1}}

\begin{document}
\thispagestyle{empty}

\def\thefootnote{\fnsymbol{footnote}}

\begin{flushright}
ANL-HEP-PR-12-84
\end{flushright}

\vspace{2cm}

\begin{center}

{\Large\sc {\bf Higgs CP Properties From Early LHC Data}}
\\[3.5em]
{\large\sc
A.~Freitas$^1$, P.~Schwaller$^{2,3}$
}

\vspace*{1cm}

{\sl $^1$
Pittsburgh Particle physics, Astrophysics \& Cosmology
    Center (PITT-PACC),  Department of Physics \& Astronomy,
    University of Pittsburgh, Pittsburgh, PA 15260, USA
}
\\[1em]
{\sl $^2$
HEP Division, Argonne National Laboratory,
 9700 Cass Ave, Argonne, IL 60439, USA}
\\[1em]
{\sl $^3$
Department of Physics, University of Illinois, 845 W Taylor St, Chicago, IL 60607, USA}

\end{center}

\vspace*{2.5cm}

\begin{abstract}
In this paper, we constrain CP violation in the Higgs
sector using the measured signal strengths in the various Higgs search channels.
To this end, we introduce a general parameterization for a resonance which is an
admixture of a CP-even Higgs-like state and a CP-odd scalar. By performing a fit to
the available data from the Tevatron and LHC experiments, one obtains constraints
on the mixing
angle and the couplings of the resonance to Standard Model fields. Depending on
the couplings, sizable mixing angles are still compatible with the data, but
small mixing is in general preferred by the fit. In particular, we find that a
pure CP-odd state is disfavored by the current data at the $3\sigma$ level.
Additionally, we consider a mixed fermiophobic resonance and a model with two
degenerate mixed resonances and find that both scenarios can successfully fit
the data within current errors. Finally, we estimate that the mixing
angle can be constrained to $\alpha < 1.1$ (0.7) in the full 8~TeV (14~TeV) run
of the LHC. 
\end{abstract}

\def\thefootnote{\fnsymbol{footnote}}
\setcounter{page}{0}
\setcounter{footnote}{0}

\newpage


\section{Introduction}

Recently, the ATLAS and CMS collaborations have reported on the discovery of a
new bosonic resonance with mass in the range 125--126~GeV at the Large Hadron
Collider (LHC) \cite{atlash,cmsh}, which has been corroborated by an excess
observed by the CDF and D\O\ experiments at the Tevatron \cite{tevh,tevh2}.
While the current data is in agreement with expectations for the Standard Model
(SM) Higgs boson, the experimental uncertainties are still large,
and thus other possibilities still need to be considered. 

Of particular interest are the spin and CP quantum numbers of the new particle.
Since it is known to decay into photon pairs, it cannot be a spin-1
particle. A spin-2 resonance may be distinguished from a spin-0 resonance by
analyzing angular distributions in the $\gamma\gamma$ \cite{spingw,spinggzg}, $ZZ^* \to
4\ell$ \cite{zzCP,zzCP8,spinzz}, $WW^* \to \ell\nu\ell\nu$ \cite{spingw}, and
$Z\gamma \to \ell\ell\gamma$ \cite{spinzg,spinggzg} decay channels, or angular
correlations in associated production with jets \cite{spinjets}. Furthermore, a
spin-0 particle $\phi$ may be CP-even, CP-odd, or a general mixed CP state. The
CP properties can be determined from angular distributions in $ZZ^* \to 4\ell$
\cite{zzCP,zzCP8}, angular distributions of the jets in $\phi+2$~jets production
\cite{jjCP}, or from spin correlations in $\phi \to \tau^+\tau^-$ decays
\cite{tautauCP}.

However, the analysis of distributions becomes viable only if a sufficient
number of events has been accumulated in a given channel. At this early stage,
however, one can already constrain the CP properties from the observed
production rates and decay branching fractions \cite{br,Coleppa:2012eh}. This mainly follows from the
fact that a CP-odd even-spin particle cannot have renormalizable tree-level
couplings to two gauge bosons.
Based on this approach,
Ref.~\cite{Coleppa:2012eh} finds that a CP-odd pseudoscalar is disfavored compared to
a CP-even scalar, although their conclusion is not based on a global fit to the
known data and is thus difficult to interpret statistically. The goal of this paper
is to carry out such a fit in a general setup where the 125-GeV resonance can
be an arbitrary mixture of CP-even and CP-odd components, and can have modified
couplings to SM fermions as well as new couplings to SM gauge bosons mediated
through higher-dimensional operators.

The model setup is explained in more detail in section~\ref{setup}. The
possibility of general CP mixing leads to modified decay branching fractions and
production rates, which are discussed in section~\ref{decprod}. In
section~\ref{num}, these observables
are then confronted with the available experimental data from July 2012, to put
constraints on the amount of CP mixing and coupling parameters. Finally,
projections for how these bounds may improve with additional data from the LHC
are presented in section~\ref{future}, before concluding in section~\ref{concl}.


\section{Setup}
\label{setup}

Throughout this paper, it will be assumed that the 125-GeV resonance observed by
ATLAS and CMS, denoted $\phi$, is a scalar, but its CP properties are left
unconstrained. In general, it can be a mixture of a CP-even Higgs-like scalar
$H$ and a CP-odd scalar $A$:
\begin{equation}
\phi = \cos\alpha\;H + \sin\alpha\;A\,. \label{hmix}
\end{equation}
CP mixing in the Higgs sector can appear in many extensions of the SM. Two of
the simplest possibilities are a complex singlet extension of the SM \cite{csm}
and the Two-Higgs-Doublet Model (THDM) \cite{thdm}\footnote{The LHC Higgs data has been analyzed in the context of specific
realizations of the THDM in several recent papers
\cite{thdm2}.}. As a result of eq.~\eqref{hmix}, the tree-level couplings of $\phi$ to
$W$ and $Z$ bosons are reduced by a factor $\cos\alpha$ compared to the SM 
since $AW^+W^-$ and $AZZ$ couplings can be realized only through operators of
dimension five or higher and thus are expected to be generated through loops of
heavy particles. 

Generically, we assume that the orthogonal state
\begin{equation}
\phi' = -\sin\alpha\;H + \cos\alpha\;A \label{hpmix}
\end{equation}
is much heavier than $\phi$ and evades current search limits through its
modified couplings compared to the SM Higgs.

In a general extension of the SM, the Yukawa couplings of the CP-even and CP-odd
components of $\phi$ are free parameters. However, existing data on the fermion
masses and mixings essentially demands that the up-type and down-type Yukawa
matrices can be written as the SM Yukawa matrices $Y^{\rm u,d,\ell}$ times some
overall constant for each matrix. This is described by the Lagrangian
\begin{equation}
\begin{aligned}
{\cal L}_{\rm Yuk} = \, & -y_{\rm u} Y^{\rm u}_{ij} \bar{u}_i u_j H
                          -y_{\rm d} Y^{\rm d}_{ij} \bar{d}_i d_j H 
			  -y_{\rm d} Y^{\ell}_{ij} \bar{\ell}_i \ell_j H \\
			& -ix_{\rm u} Y^{\rm u}_{ij} \bar{u}_i u_j A
                          -ix_{\rm d} Y^{\rm d}_{ij} \bar{d}_i d_j A 
			  -ix_{\rm d} Y^{\ell}_{ij} \bar{\ell}_i \ell_j A
			+ \text{h.c.}\,,
\end{aligned} \label{yuk}
\end{equation}
where $y_{\rm u,d}$ and $x_{\rm u,d}$ parametrize the strength of the CP-even
and CP-odd Yukawa couplings, respectively, relative to the SM coupling strength.
In particular, the THDM types I and II fit in this pattern. 
Note that the framework in eqs.~\eqref{hmix} and \eqref{yuk} is general
enough to accommodate the possibility that $H$ itself is a mixture of several
CP-even states---in this case $\alpha$, $y_{\rm u}$ and $y_{\rm d}$ would be functions
of the $3\times 3$ Higgs mixing matrix. The SM corresponds to the choices $\alpha=0$,
$y_{\rm u,d}=1$, and $x_{\rm u,d}=0$. 

While there are no renormalizable couplings of the CP-odd component $A$ to the
SM gauge bosons, higher-dimensional interaction operators may be induced through
loop corrections of heavy new fields. In an effective field
theory formulation these interactions are given by
\begin{align}
 {\cal L}_{\rm dim5} = 
  \frac{1}{4} \frac{c_G}{(4 \pi)^2v} A G_{\mu\nu}\widetilde{G}^{\mu\nu} + 
  \frac{1}{4}\frac{c_B}{(4\pi)^2v} A B_{\mu\nu} \widetilde{B}^{\mu\nu} + 
  \frac{1}{4} \frac{c_W}{(4\pi)^2v} A W_{\mu\nu} \widetilde{W}^{\mu\nu}\,, 
  \label{dim5}
\end{align}
with $\widetilde{G}^{\mu\nu} = \epsilon^{\mu\nu\alpha\beta}G_{\alpha\beta}$
\emph{etc}., and $v=174\gev$ is the electroweak vacuum expectation value. 
The normalization is chosen such that the Feynman rules have a 
prefactor $c_i/(16 \pi^2 v)$ (see appendix). 
We assume that the coefficients originate from new
perturbative physics, so that $c_i < 4\pi$. 
We do not consider dimension-five operators for the coupling of the CP-even
component $H$ to gauge bosons 
since the effects are typically small compared to the
tree-level $HWW$ and $HZZ$ couplings, while the loop-induced $H\gamma\gamma$ and
$Hgg$ interactions can be sufficiently generally described by the modified
Yukawa couplings in eq.~\eqref{yuk}.


\section{Decay Widths and Production Rates}
\label{decprod}

Let us begin by disregarding the dimension-5 operators in eq.~\eqref{dim5} in order to 
illustrate the effect of the CP mixing and modified Yukawa couplings. Compared
to the SM, the partial widths for the tree-level decays are given by
\begin{align}
&\frac{\Gamma[\phi \to WW^*]}{\Gamma_{\rm SM}[H\to WW^*]}
= \frac{\Gamma[\phi \to ZZ^*]}{\Gamma_{\rm SM}[H\to ZZ^*]} = \cos^2\alpha\,,
\label{GVV}
\\[.5ex]
&\frac{\Gamma[\phi \to \tau^+\tau^-]}{\Gamma_{\rm SM}[H\to \tau^+\tau^-]}
= (y_{\rm d}\cos\alpha)^2 + (x_{\rm d}\sin\alpha)^2\,,
\\[.5ex]
&\frac{\Gamma[\phi \to c\bar{c}]}{\Gamma_{\rm SM}[H\to c\bar{c}]}
= (y_{\rm u}\cos\alpha)^2 + R^{c\bar{c}}(x_{\rm u}\sin\alpha)^2\,,
\\[.5ex]
&\frac{\Gamma[\phi \to b\bar{b}]}{\Gamma_{\rm SM}[H\to b\bar{b}]}
= (y_{\rm d}\cos\alpha)^2 + R^{b\bar{b}}(x_{\rm d}\sin\alpha)^2\,, \label{eqn:Gbb}
\intertext{while the loop-induced decay widths read \cite{loop}}
&\frac{\Gamma[\phi \to gg]}{\Gamma_{\rm SM}[H\to gg]}
= \cos^2\alpha \,\frac{|y_{\rm u} H_{1/2}(\tau_\Pt) + y_{\rm d}
 H_{1/2}(\tau_\Pb)|^2}{|H_{1/2}(\tau_\Pt)|^2} +
 \sin^2\alpha \,R^{gg}\frac{|x_{\rm u} A_{1/2}(\tau_\Pt) + x_{\rm d}
 A_{1/2}(\tau_\Pb)|^2}{|H_{1/2}(\tau_\Pt)|^2}\,, \label{Ggg} \\[1ex]
&\frac{\Gamma[\phi \to \gamma\gamma]}{\Gamma_{\rm SM}[H\to \gamma\gamma]}
= \begin{aligned}[t]
 &\cos^2\alpha \,\frac{|\frac{4}{3}y_{\rm u} H_{1/2}(\tau_\Pt) + \frac{1}{3}y_{\rm d}
 H_{1/2}(\tau_\Pb) + y_{\rm d} H_{1/2}(\tau_\tau) -
 H_1(\tau_\PW)|^2}{|\frac{4}{3} H_{1/2}(\tau_\Pt) - H_1(\tau_\PW)|^2} \\ &+
 \sin^2\alpha \,\frac{|\frac{4}{3}x_{\rm u} A_{1/2}(\tau_\Pt) + \frac{1}{3}x_{\rm d}
 A_{1/2}(\tau_\Pb) + x_{\rm d} A_{1/2}(\tau_\tau)|^2}{|\frac{4}{3} H_{1/2}(\tau_\Pt) -
 H_1(\tau_\PW)|^2}\,,
\end{aligned} \label{Gaa}
\end{align}
where $\tau_f = m_\phi^2/(4 m_f^2)$, and
\begin{align}
H_{1/2}(\tau) &= \frac{(\tau-1)f(\tau)+\tau}{\tau^2}\,, \quad
H_1(\tau) = \frac{3(2\tau-1)f(\tau) + 3\tau + 2\tau^2}{2\tau^2}\,, \quad
A_{1/2}(\tau) = \frac{f(\tau)}{\tau}\,, \\[1ex]
f(\tau) &= \left\{\begin{array}{ll}
 \arcsin^2(\sqrt{\tau}) & (\tau \leq 1)\,, \\[.5ex]
 -\frac{1}{4}\Bigl ( \log\frac{1+\sqrt{1-1/\tau}}{1-\sqrt{1-1/\tau}}-i\pi
  \Bigr )^2 & (\tau > 1)\,.
\end{array}\right.
\end{align}
For the SM decay rates into $gg$ and $\gamma\gamma$ in \eqref{Ggg} and
\eqref{Gaa}, the
contributions from tau leptons and bottom quarks may be safely neglected, but
for $\phi$ decays they can be enhanced by large Yukawa factors $y_{\rm d}$ and
$x_{\rm d}$ and thus need to be included.
The factors $R^X$ incorporate the difference between the QCD corrections for
scalar and pseudoscalar decays, $R^X = \frac{1+\Delta_{\rm QCD}[A\to
X]}{1+\Delta_{\rm QCD}[H\to X]}$ (for a review see Ref.~\cite{qcd}). 
They deviate from unity by less than 1\%.

The production rates at the Tevatron and LHC for final states
$X=WW^*,ZZ^*,\gamma\gamma,\tau\tau$ can then be written as
\begin{equation}
\begin{aligned}
r_X &\equiv \frac{\sigma[p\,p\rput[rb](.3ex,1ex){\psscaleboxto(1.8ex,1ex){\bf (-)}}\, \to \phi \to X]}%
{\sigma_{\rm SM} [p\,p\rput[rb](.3ex,1ex){\psscaleboxto(1.8ex,1ex){\bf (-)}}\, \to H \to X]} 
\\
&= \biggl ( f_{gg} \frac{\Gamma[\phi \to gg]}{\Gamma_{\rm SM}[H \to gg]} +
           f_{\rm VBF} \frac{\Gamma[\phi \to WW^*]}{\Gamma_{\rm SM}[H \to WW^*]}
	   \biggr ) \times
 \frac{\Gamma^{\rm SM}_{H,\rm tot}}{\Gamma_{\phi,\rm tot}} \times
 \frac{\Gamma[\phi \to X]}{\Gamma_{\rm SM}[H \to X]}\,.
\end{aligned}
\end{equation}
Here $f_{gg}$ and $f_{\rm VBF}$ denote the fractions with which the two
dominant production modes, gluon fusion and vector-boson fusion, contribute to
the total production cross section, respectively ($f_{gg} + f_{\rm VBF} =1$).
$\Gamma_{\phi,\rm tot}$ is the total decay width of $\phi$, which is given by
\begin{equation}
\begin{aligned}
\Gamma_{\phi,\rm tot} \approx {}&\Gamma[\phi \to WW^*] + \Gamma[\phi\to ZZ^*]
+ \Gamma[\phi \to b\bar{b}] + \Gamma[\phi \to c\bar{c}] + \Gamma[\phi \to
\tau^+\tau^-] \\ &+ \Gamma[\phi \to gg]
\end{aligned}
\end{equation}
to very good approximation\footnote{We do not consider possible non-standard
decay channels of $\phi$ in this paper.} (similar for $\Gamma^{\rm
SM}_{H,\rm tot}$). For the SM partial widths we take the values from
Ref.~\cite{tevh}. 

For the $b\bar{b}$ final state, the two leading production modes are not
experimentally viable due to large backgrounds. Instead, the experimental
collaborations focus on associated production with a $W$ or $Z$
gauge boson, which scales with $\cos^2\alpha$ according to eq.~\eqref{GVV}. Thus
\begin{equation}
r_{b\bar{b}} = \cos^2\alpha \times
 \frac{\Gamma^{\rm SM}_{H,\rm tot}}{\Gamma_{\phi,\rm tot}} \times
 \frac{\Gamma[\phi \to b\bar{b}]}{\Gamma_{\rm SM}[H \to b\bar{b}]}\,.
\end{equation}

\vspace{\medskipamount}
If the contributions from new-physics induced higher-dimensional operators in
eq.~\eqref{dim5} are sizable, they lead to additional contributions to the
partial widths into gauge-boson pairs. The relevant Feynman rules are listed in
the appendix. Including these terms, one obtains (with the approximation $R^{gg}
\approx 1$)
\begin{align}
&\begin{aligned}
\frac{\Gamma[\phi \to gg]}{\Gamma_{\rm SM}[H\to gg]}
= \frac{1}{|H_{1/2}(\tau_\Pt)|^2} &\Bigl ( 
 \cos^2\alpha \, \bigl|y_{\rm u} H_{1/2}(\tau_\Pt) + y_{\rm d}
 H_{1/2}(\tau_\Pb)\bigr|^2 \\ 
 &+ \sin^2\alpha \,\bigl|x_{\rm u} A_{1/2}(\tau_\Pt) + x_{\rm d}
 A_{1/2}(\tau_\Pb) + \sqrt{2}c_G/g_{\rm s}^2 \bigr|^2 \Bigr ) \,,
 \label{eqn:gg5}
\end{aligned}
\\
&\begin{aligned}
&\frac{\Gamma[\phi \to \gamma\gamma]}{\Gamma_{\rm SM}[H\to \gamma\gamma]} 
= \frac{1}{|\frac{4}{3} H_{1/2}(\tau_\Pt) - H_1(\tau_\PW)|^2} \\
 &\qquad\qquad \times \biggl ( 
 \cos^2\alpha \,\bigl|\tfrac{4}{3}y_{\rm u} H_{1/2}(\tau_\Pt) + \tfrac{1}{3}y_{\rm d}
 H_{1/2}(\tau_\Pb) + y_{\rm d} H_{1/2}(\tau_\tau) -
 H_1(\tau_\PW) \bigr|^2 \\ 
 &\qquad\qquad\quad\;+
 \sin^2\alpha \,\biggl| \tfrac{4}{3}x_{\rm u} A_{1/2}(\tau_\Pt) + \tfrac{1}{3}x_{\rm d}
 A_{1/2}(\tau_\Pb) + x_{\rm d} A_{1/2}(\tau_\tau) 
 + \frac{(c_\theta^2c_B+s_\theta^2c_W)}{\sqrt{2}e^2} \biggr|^2 \biggr ) \,,
\end{aligned} 
\end{align}
where $c_\theta \equiv  \cos\theta_{\rm W}$,  $s_\theta \equiv 
\sin\theta_{\rm W}$, and $\theta_{\rm W}$ is the Weinberg angle. 
For the $\gamma Z$ decay one arrives at a similar expression, which we do not
write down here since it is rather lengthy \cite{zgamma}. In fact, the SM contribution to this
decay channel is rather small and thus irrelevant for the current early stage of
Higgs searches. However, the dimension-5 operators in eq.~\eqref{dim5} could
potentially lead to a much larger result, that would dominate over the SM
contribution, in which case one can write
\begin{equation}
\Gamma[\phi \to \gamma Z] \approx \sin^2\alpha \,
 \frac{s_{2\theta}^2(c_W-c_B)^2}{8(4\pi)^5v^2}\;
 \frac{(m_\phi^2-\MZ^2)^3}{m_\phi^3}\,.
\end{equation}
For the four-body decay modes mediated by $WW$ and $ZZ$ pairs one finds, using
{\sc CalcHEP}~\cite{calchep} for the numerical phase-space integration,
\begin{align}
&\frac{\Gamma[\phi \to WW^*]}{\Gamma_{\rm SM}[H\to WW^*]} =
\cos^2\alpha + \sin^2 \alpha \frac{c_W^2}{(4\pi)^4} \times 0.155 \,, \\
&\frac{\Gamma[\phi \to ZZ^*]}{\Gamma_{\rm SM}[H\to ZZ^*]} = 
\cos^2\alpha + \sin^2 \alpha  \frac{(s_\theta^2c_B + c_\theta^2c_W)^2}{(4\pi)^4} \times 0.074 \,.
\label{eqn:zz5}
\end{align}
Finally, it is important to note that there are no interference effects between
the CP-even and CP-odd contributions in the inclusive rates, in contrast to
specifically CP-sensitive observables such as certain angular distributions
\cite{zzCP,zzCP8,jjCP,tautauCP}.


\section{Numerical Analysis of Summer 2012 Data}
\label{num}

In this section, the formalism of the previous two sections is applied to the
experimental Higgs search results released in July 2012 by the Tevatron and LHC
collaborations
\cite{tevh,atlash,cmsh,atlasvbf,cmsvbf}. It is shown that the CP properties of the
new resonance can already be constrained substantially from our current
knowledge of its production rates and branching fractions, even though the
experimental uncertainties are still large. The relevant input data, as read off
from the plots in Refs.~\cite{tevh,atlash,cmsh,atlasvbf,cmsvbf}, is summarized in
Tab.~\ref{tab:measured}. The following values are taken for the relative
production rates:
\begin{itemize}
\item Tevatron inclusive: $f_{gg} \approx 0.78$, $f_{\rm VBF} \approx 0.22$ \cite{tevh};
\item LHC inclusive: $f_{gg} \approx 0.9$, $f_{\rm VBF} \approx 0.1$
 \cite{atlasvbf};
\item LHC $\gamma\gamma$ VBF enhanced: $f_{gg} \approx 0.25$, $f_{\rm VBF} \approx 0.75$
 \cite{atlasvbf};
\item LHC $\tau\tau$ analysis: $f_{gg} \approx 0.5$, $f_{\rm VBF} \approx 0.5$;
\newline
estimated from the observation that inclusive and VBF-enhanced measurements of
this channel contribute with approximately similar significance \cite{tautausub}.
\end{itemize}
Here ``VBF enhanced'' refers to Higgs searches with a set of cuts that enhance
the relative contribution of the vector-boson fusion production mode, 
characterized by two energetic jets with a large rapidity gap. Since both these
searches and the inclusive LHC measurements receive contributions from gluon
fusion and VBF, there is some degree of correlation between
them, which is taken into account with a covariance matrix in the $\chi^2$ fit.

\begin{table}[t]
\centering
\renewcommand{\arraystretch}{1.3}
\begin{tabular}{|l|l|l|c|}
\hline
Experiment & $X$ & $r_X^{}/r^{\rm SM}_X$ & Ref. \\
\hline\hline 
\multirow{7}{*}{ATLAS} & $WW^*$ & $1.24 \pm 0.45$ & \\
\cline{2-3}
 & $ZZ^*$ & $1.39 \pm 0.60$ & \\
\cline{2-3}
 & $b\bar{b}$ & $0.50^{+2.13}_{-2.18}$ & \cite{atlash} \\
\cline{2-3}
 & $\tau\tau$ & $0.45^{+1.54}_{-2.04}$ & \\
\cline{2-3}
 & $\gamma\gamma$ & $1.79 \pm 0.50$ & \\
\cline{2-4}
 & $\gamma\gamma$ (VBF enh.\ 7 TeV) & $4.19 \pm 2.10$ &
 \multirow{2}{*}{\hspace{0.0001ex}\cite{atlasvbf}} \\
\cline{2-3}
 & $\gamma\gamma$ (VBF enh.\ 8 TeV) & $1.24 \pm 1.57$ & \\
\hline
\multirow{6}{*}{CMS} & $WW^*$ & $0.59^{+0.46}_{-0.38}$ & \\
\cline{2-3}
 & $ZZ^*$ & $0.72^{+0.48}_{-0.35}$ & \\
\cline{2-3}
 & $b\bar{b}$ & $0.48^{+0.83}_{-0.72}$ & \cite{cmsh} \\
\cline{2-3}
 & $\tau\tau$ & $0.08^{+0.81}_{-0.75}$ & \\
\cline{2-3}
 & $\gamma\gamma$ & $1.56 \pm 0.47$ & \\
\cline{2-4}
 & $\gamma\gamma$ (VBF enh.) & $2.30 \pm 1.26$ & \cite{cmsvbf} \\
\hline
 & $WW^*$ & $0.32^{+1.13}_{-0.32}$ & \\
\cline{2-3}
CDF/D\O & $b\bar{b}$ & $1.97^{+0.74}_{-0.68}$ & \cite{tevh} \\
\cline{2-3}
 & $\gamma\gamma$ & $3.62^{+2.96}_{-2.54}$ & \\
\hline
\end{tabular}
\mycaption{Experimental results for Higgs production rates in different
final-state channels from Tevatron and LHC used in this analysis. Separately
shown are the values for $\gamma\gamma$ final states with cuts to enhance the
VBF production mode from ATLAS and CMS.}
\label{tab:measured}
\end{table}
%
%
Fitting this data to the SM predictions we find 
\begin{align}
	\chi^2_{\rm SM} = 13.3 \quad (\text{16 d.o.f.})\,.
\end{align}
For a $\chi^2$ distribution with 16 degrees of freedom, the $68\%$ $(95\%)$
confidence limit (C.L.) corresponds to $\chi^2=17.0$ $(25.0)$. Thus one can see that
the overall agreement of the data with the SM prediction is very good.  For the
two parameter plots that will be shown later, the $68\%$, $95\%$, and $99.7\%$
contours correspond to $\chi^2 = 15.9$, $23.7$, and $32.9$, respectively. 


\subsection{Single Resonance}
\label{1res}

We first consider the scenario described by equations~(\ref{hmix})
and~(\ref{yuk}). Here both $H$ and $A$ have renormalizable couplings to the SM
fermions, while the couplings of $A$ to SM gauge bosons are induced at the one
loop level through SM fermion loops. We assume that there are no other new
physics states that generate couplings of $A$ to SM gauge bosons, and therefore
$c_G = c_B = c_W = 0$ in~(\ref{dim5}).  The fermionic couplings of $H$ and $A$ are
allowed to deviate from their SM values through the parameters $y_{u,d}$ and
$x_{u,d}$, respectively. As mentioned in section~\ref{setup}, 
it is assumed that $m_\phi=125\gev \ll m_{\phi'}$. 

Let us first consider the case where the CP-even state has SM-like Yukawa
couplings, $y_{\rm u} = y_{\rm d} = 1$, while the CP-odd couplings $x_{u,d}$ are
unknown. In the limit of zero CP-odd couplings, all channels are uniformly
suppressed by $\cos^2\alpha$. For this particular point we find 
\begin{align}
	\alpha < 0.76 \qquad (95\% \text{ C.L.})\,,
\end{align}
while the best fit is for $\alpha=0$. 
When allowing the CP-odd couplings to float freely, the overall rate suppression
from the mixing can now be offset with an increase in the production rate when
$x_{\rm u}> 0$. In fact, large values of $x_{\rm u}$ are favored in the fit, so
to ensure perturbativity of the top Yukawa coupling we impose an upper limit
$x_{\rm u} < 3$. 
The effects of $x_{\rm d}$ are more subtle. It can increase the
total width and thus suppress all but the $\tau\tau$ and $b\bar{b}$ channels, so that 
large values of $x_d$ do not produce a good fit. Overall, we find that a
marginally better $\chi^2$ than for the SM is obtained for nonzero but small
mixing $\alpha = 0.07$, maximal $x_{\rm u}=3$, and vanishing $x_{\rm d}$. 
%
%
%
\begin{figure}[t]
\epsfig{width=0.45\textwidth, file=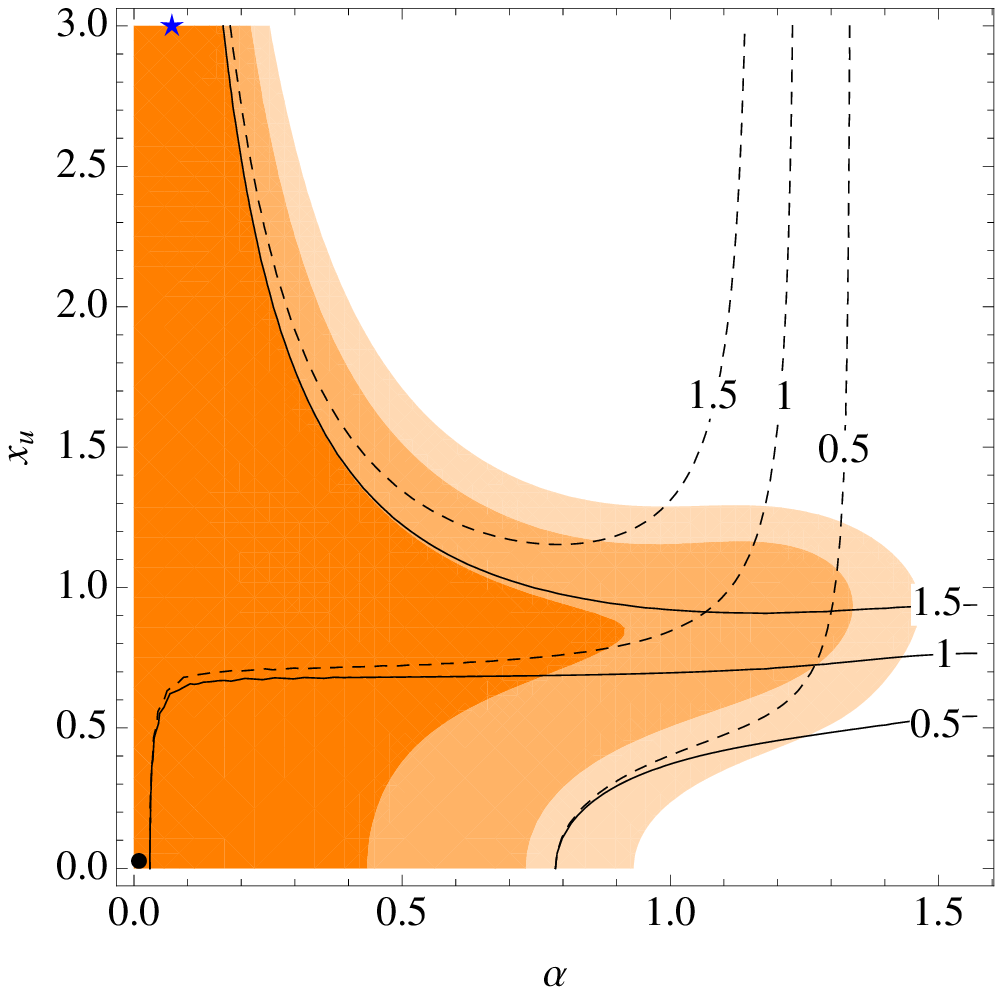}\hfill
\parbox[b]{0.49\textwidth}{
\mycaption{Quality of the fit to experimental data in the $x_{\rm u}-\alpha$
plane, for $x_{\rm d}=0$ and $y_{\rm u}=y_{\rm d}=1$. The orange (grey) shaded
areas agree with the data at the $1\sigma$ (dark), $2\sigma$ (medium) and
$3\sigma$ (light) level. The blue star shows the best fit point, while the black
dot corresponds to the SM. The solid (dashed) lines are contours of constant
$r_{\gamma \gamma}$ $(r_{ZZ})$.
\newline}
\label{fig:res1}}
\end{figure}

The large value of $x_{\rm u}$ together with a small mixing leads to slightly
enhanced signal rates at the 10\% level across all channels, which is slightly
favored by the current data. The overall quality of the fit in the $x_{\rm
u}$--$\alpha$ plane for $x_{\rm d} = 0$ is shown in Fig.~\ref{fig:res1}. Mixing angles of up to $\alpha = 1.3$ are compatible with the data at the
95\% C.L. 

Close to $\alpha=\pi/2$, the field $\phi$ becomes mostly CP-odd, and the signal
rates $r_{ZZ}$ and $r_{WW}$ become strongly suppressed. For smaller mixing, both
$r_{ZZ}$ and $r_{\gamma\gamma}$ can be enhanced or reduced relative to the SM. 
However, an enhancement of the di-photon rate by more than 50\% is only possible outside
of the $1\sigma$ region. 

\

Let us now consider the case where the CP-even Yukawa couplings can
vary with respect to the SM. Due to the additional free parameters, the
predicted rates in the different channels are less strongly correlated with each
other. For nonzero
mixing, there is some redundancy in the couplings $x_{\rm u}$--$y_{\rm u}$ and
$x_{\rm d}$--$y_{\rm d}$, which leads to almost flat $\chi^2$ distributions in some
directions. 

Numerically, we find the minimum of the $\chi^2$ distribution at
\begin{align}
	\alpha = 0.16\,, \quad x_{\rm u} = 3.0\,, \quad x_{\rm d} = y_{\rm u} = 0 \,, \quad y_{\rm d}=0.74\,. \label{eqn:bestfit}
\end{align}
With $\chi^2 = 7.6$, this parameter point lies significantly below the SM fit,
however at the expense of having five additional model parameters. Both the
CP-mixing scenario and the SM have $\chi^2/({\rm d.o.f.})<1$, $i.\,e.$ the
current data does not conclusively favor one model over the other.
Nevertheless, we will indicate the most preferred regions with $\Delta \chi^2<1$ from the
minimum in the plots. Should future (more precise) data have similar central
values, those regions would be strongly favored. 

The good quality of the fit can be understood as follows. First, the gluon
fusion production is reduced to roughly 65\% compared to the SM, which is mostly
due to the vanishing CP-even Yukawa coupling, $y_{\rm u}=0$, while the small mixing angle suppresses the coupling of
the CP-odd component to gluons. The mixing also slightly suppresses the VBF
channel to about 85\%. The di-photon decay width is naturally enhanced
since the destructive interference between the top and the $W$-boson loop goes
away. Additional enhancement comes from the large decay width of the CP-odd
component into photons, and from a reduction of the dominant 
$b\bar{b}$ width due to $y_{\rm d} <1$. The
change in the $WW$ and $ZZ$ channels is more balanced, since the increased
branching ratio to those final states is compensated by the overall reduced
production cross section. Finally, $r_{\tau\tau}$ is suppressed by about 50\% due
to the smallness of $y_{\rm d}$. 

\begin{figure}[tp]
\begin{center}
\epsfig{width=0.45\textwidth, file=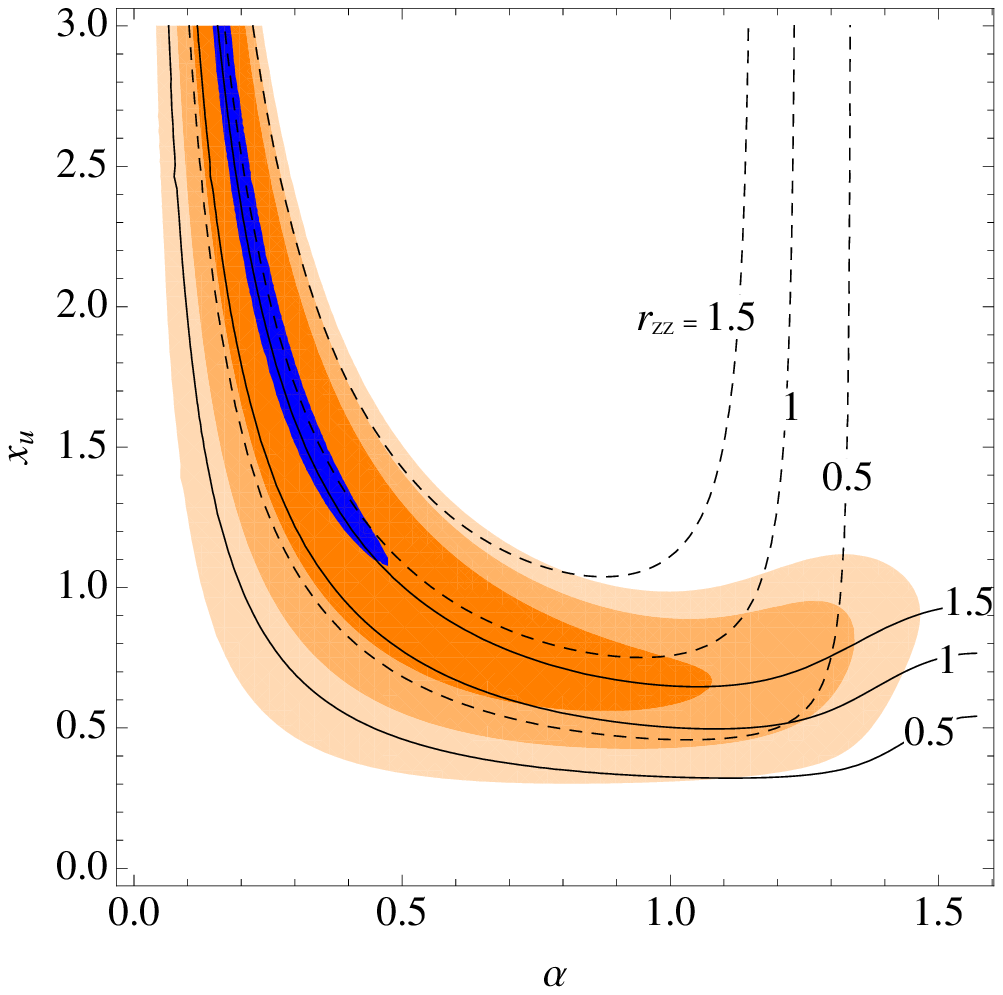}\hspace{1cm}
\epsfig{width=0.45\textwidth, file=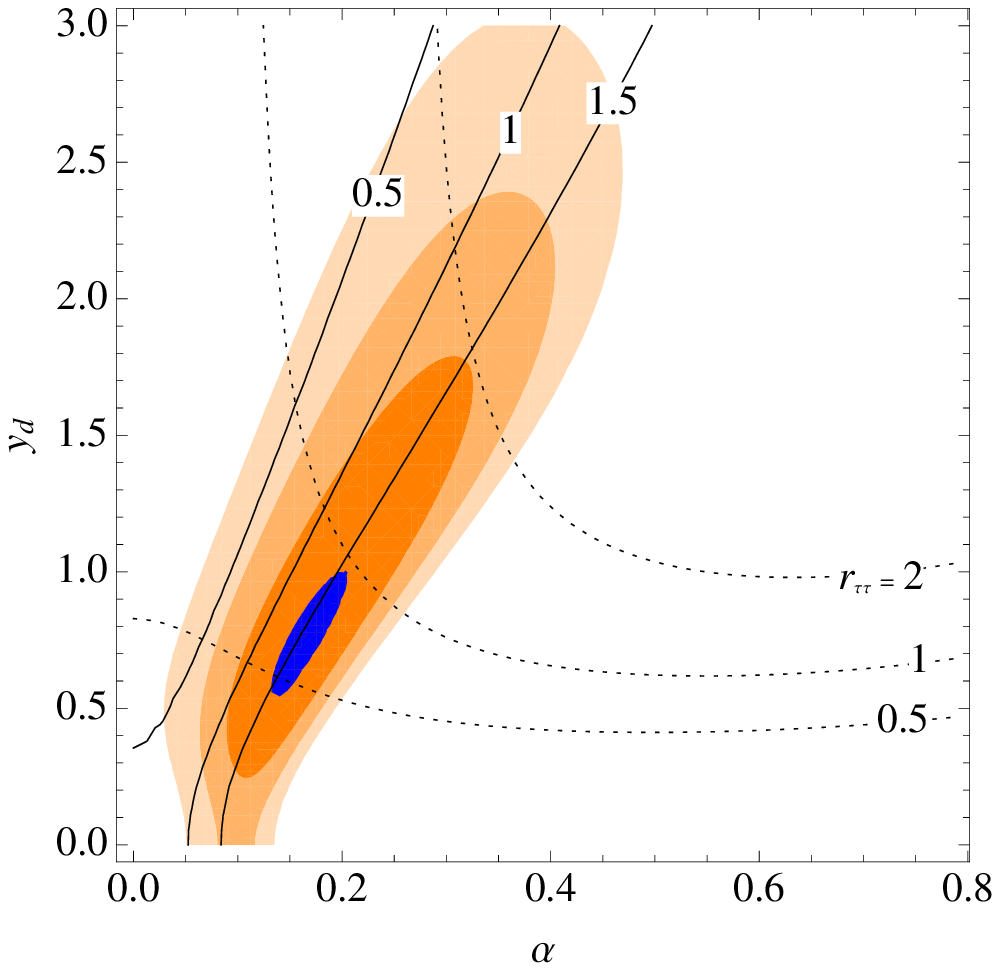}\\[.5cm]
\epsfig{width=0.45\textwidth, file=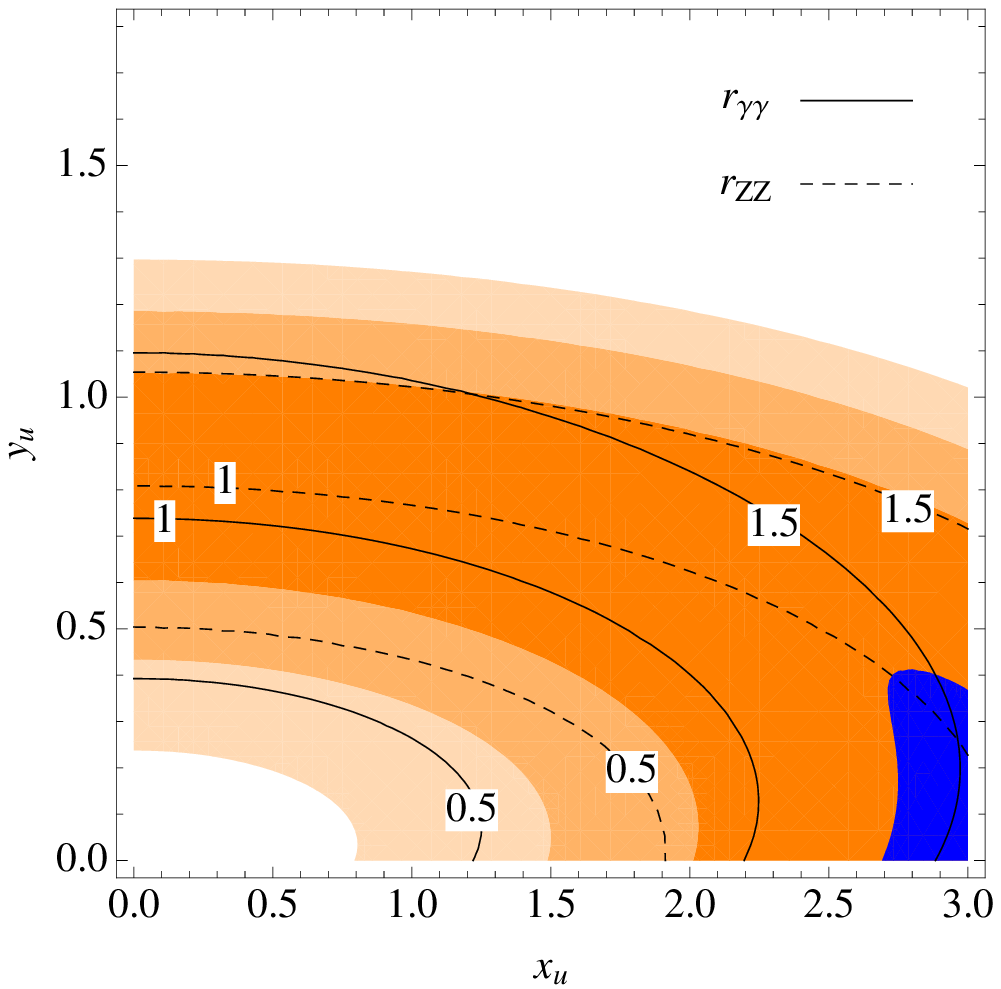}\hspace{1cm}
\epsfig{width=0.45\textwidth, file=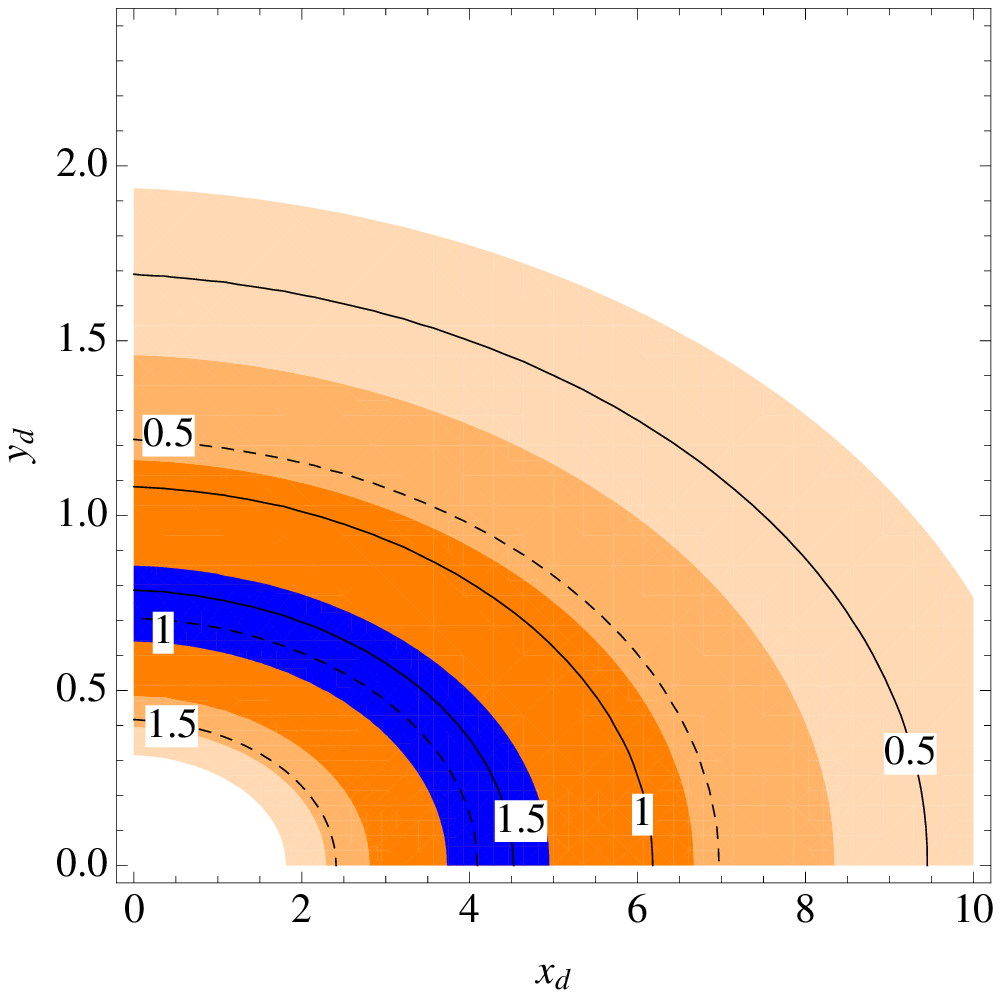}
\end{center}
\vspace{-1em}
\mycaption{Quality of the fit for the single resonance model in the
$\alpha$--$x_{\rm u}$ (top left), $\alpha$--$y_{\rm d}$ (top-right), $x_{\rm u}$--$y_{\rm
u}$ (bottom left), and $x_{\rm d}$--$y_{\rm d}$ (bottom right) plane. The remaining
free parameters are set to their best fit values (see text for details). Colors
and contours are as in Fig.~\ref{fig:res1}.  In addition, the blue (very dark) shaded
region indicates $\Delta \chi^2<1$ relative to the best fit, and the dotted
lines in the top-right plot are contours of constant $r_{\tau\tau}$.}
\label{fig:res2}
\end{figure}

The dependence of the fit on the different model parameters is illustrated in
Fig.~\ref{fig:res2}. The various plots show a scan over two parameters, while
the remaining free parameters are set to the best-fit values in
(\ref{eqn:bestfit}).  We can easily see that the best-fit regions typically have
$r_{\gamma\gamma} \approx 1.5$, while the $ZZ$ and $WW$ rates are kept closer to
one. 

The correlation between $x_{\rm u}$ and the mixing angle is very strong. When
$x_{\rm u}$ is reduced, a larger mixing is required to fit the data since 
otherwise the total production cross section becomes too small. Similar to
Fig.~\ref{fig:res1}, the $ZZ$ and $WW$ channels are strongly suppressed for
$\alpha \gesim 1.0$, so that this region never leads to a satisfactory fit.

The interplay between $y_{\rm d}$ and $\alpha$ in the top right plot of
Fig.~\ref{fig:res2} is again more subtle. Since $x_{\rm u}$ is fixed and $y_{\rm
u}=0$ here, increasing $y_{\rm d}$ leads to a suppression of the gauge boson
channels as the total width goes up. To some extent this can be compensated
with an increase in $\alpha$, which increases the total production cross section.
Eventually, this leads to a strong enhancement of $r_{\tau\tau}$ such that
the regions above $y_{\rm d} \approx 2.5$ are excluded here. 

The last two plots illustrate the redundancy in the couplings $x_{\rm
u}$--$y_{\rm u}$ and $x_{\rm d}$--$y_{\rm d}$. In the bottom left plot of
Fig.~\ref{fig:res2}, we see that within the $1\sigma$ contour one can trade
$x_{\rm u}$ for $y_{\rm u}$, with the ratio of the two couplings roughly given
by the mixing angle. The preference for smaller $y_{\rm u}$ comes mostly from
$r_{\gamma\gamma}$, since $y_{\rm u} < 1$ reduces the destructive interference
and thus increases the decay rate of the CP-even component into photon pairs.
Thus the redundancy between $x_{\rm u}$ and $y_{\rm u}$ can eventually be broken
with more precise data on $r_{\gamma\gamma}$.
The quarter-circle that is described by the fit contours in the $x_{\rm
d}-y_{\rm d}$ plane can easily be understood from~(\ref{eqn:Gbb}): it
corresponds to contours of constant $b\bar{b}$ width. These couplings are only
weakly constrained from the $\phi\gamma\gamma$ and $\phi g g$ couplings due to
the smallness of the bottom Yukawa coupling, so that this redundancy is
difficult to resolve in general.


\subsection{Two Near-degenerate Resonances}
\label{2res}

As pointed out above, large CP mixing can be realized in models with several
scalar multiplets. However, in the context of concrete models there are tight
constraints on CP mixing in the Higgs sector from electric dipole moments
\cite{McKeen:2012av}. It was shown in Ref.~\cite{McKeen:2012av} that
these bounds are substantially relaxed for a near-degenerate Higgs spectrum,
$i.\,e.$ if the two orthogonal states $\phi$ and $\phi'$ have almost equal
masses, $m_\phi \approx m_{\phi'}$.

If $|m_\phi - m_{\phi'}| \lesim 1\gev$, both states would contribute to the
resonance observed by ATLAS and CMS. This scenario is explored in more detail in
this subsection.\footnote{A similar study, but for Higgs mixing with a CP-even
resonance, can be found in Ref.~\cite{higgsmix}.}
 The relevant branching fractions and production rates for
$\phi'$ can be derived from the formulae in section~\ref{decprod} by making the
appropriate replacements of the mixing angles. The observed rates are then
given by
\begin{equation}
r_X = \frac{\sigma[p\,p\rput[rb](.3ex,1ex){\psscaleboxto(1.8ex,1ex){\bf
(-)}}\, \to \phi \to X] + \sigma[p\,p\rput[rb](.3ex,1ex){\psscaleboxto(1.8ex,1ex){\bf
(-)}}\, \to \phi' \to X]}%
{\sigma_{\rm SM} [p\,p\rput[rb](.3ex,1ex){\psscaleboxto(1.8ex,1ex){\bf (-)}}\, \to H \to X]} 
\,.
\end{equation}
The signal rates for $\phi'$ are obtained from the $\phi$ rates by the shift 
$\alpha \to \pi/2 -\alpha$. It therefore follows that the combined rates $r_X$
are symmetric under this transformation. For definiteness, we choose $\alpha <
\pi/4$ when searching for the best-fit point. 

Letting all model parameters float freely, we find the minimum of the $\chi^2$
distribution at 
\begin{align}
	\alpha = 0.38\,,\quad x_d = y_u = 0\,,\quad x_u = 0.57\,,\quad y_d = 0.75\,.
\end{align}
The quality of the fit is marginally better than in the single resonance model.
In both scenarios, the di-photon channel is enhanced for $y_u=0$ and $x_u>0$ 
due to absence of destructive interference as explained in section \ref{1res},
resulting in $r_{\gamma\gamma} \approx 1.6$ for the best-fit point. In addition
to the enhancement of $r_{\gamma\gamma}$ as before, now there is also a relative
enhancement of the VBF channel compared to gluon fusion production because
of the contributions of both $\phi$ and $\phi'$ exchange, such that we obtain
$r_{\gamma\gamma,VBF} > 2$. Furthermore, there is a slight suppression of
$r_{ZZ}$ and $r_{WW}$ from the mixing, and a stronger suppression of
$r_{\tau\tau}$ due to $y_d <1$.  Altogether this leads to a very good fit to the
data. Comparing with the single resonance model, the preferred value for $x_u$
is now much smaller and the mixing angle is increased, while $x_d$ and $y_{u,d}$
are roughly the same. 

\begin{figure}
\begin{center}
\epsfig{width=0.45\textwidth, file=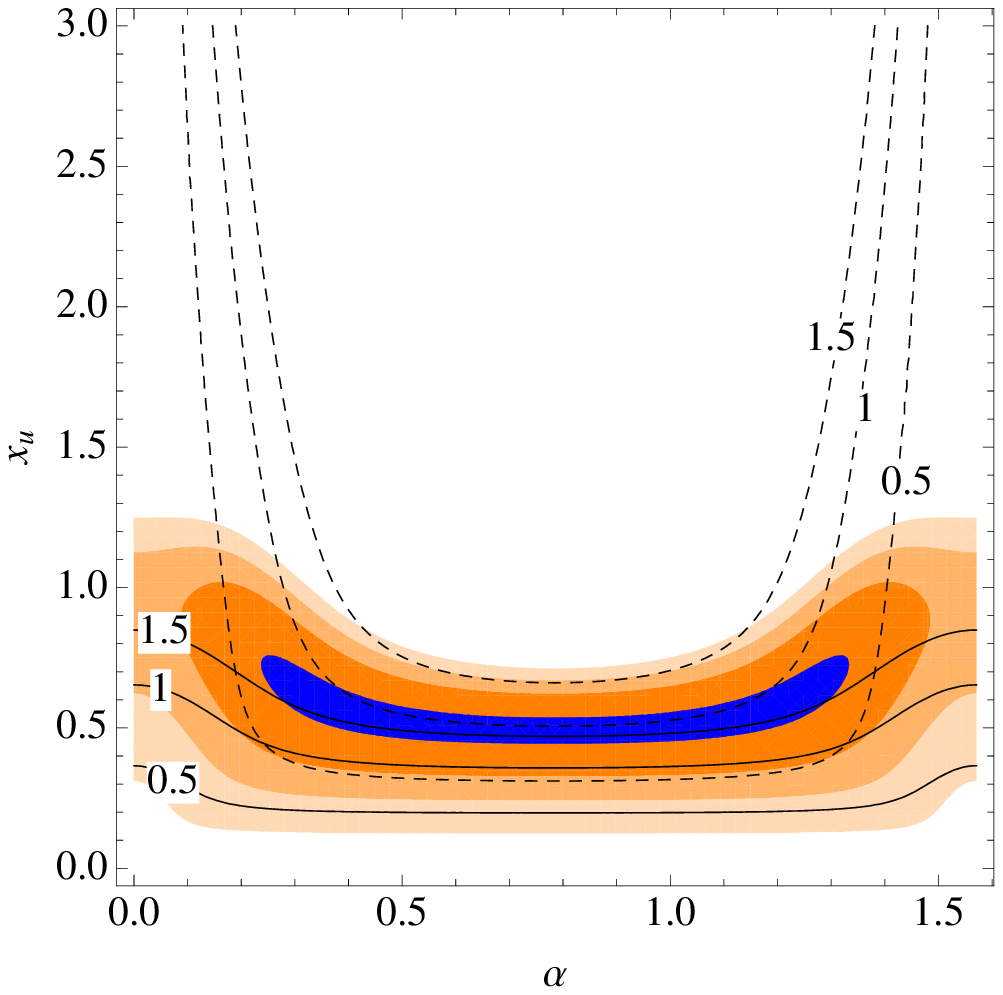}\hspace{1cm}
\epsfig{width=0.45\textwidth, file=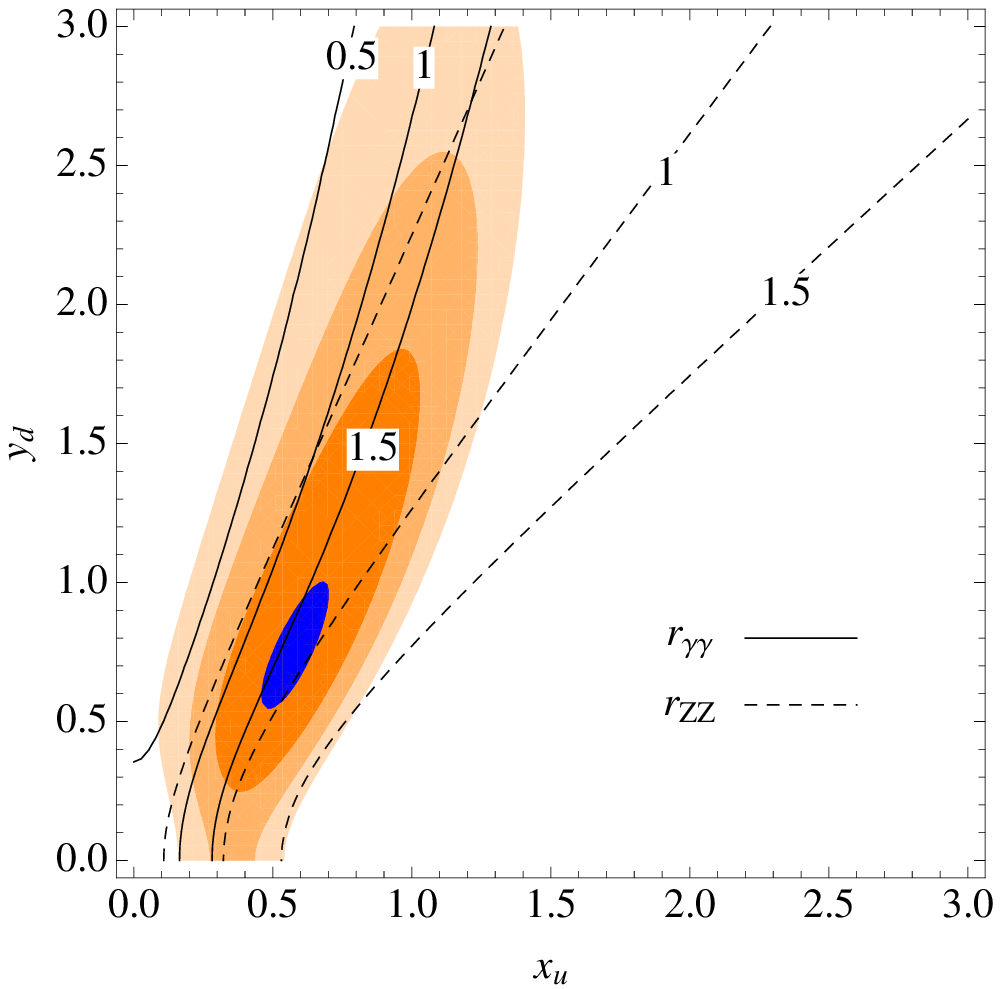}
\end{center}
\vspace{-1em}
\mycaption{Quality of the fit for the double-resonance model, in the
$\alpha$--$x_{\rm u}$ (left) and  $x_{\rm u}$--$y_{\rm d}$ (right) plane. 
The remaining free parameters are set to their best-fit values (see text for 
details). Colors and contours are as in Fig.~\ref{fig:res1} and~\ref{fig:res2}. 
}
\label{fig:res3}
\end{figure}

%
The parameter dependence of the fit in
the two-resonance model is illustrated in Fig.~\ref{fig:res3}. Compared to the model with a
single mixed resonance, there are some marked differences. First we note from
the left plot that the degeneracy between $\alpha$ and $x_u$ is now broken, and
$x_u \lesssim 1.2$ at 95\% C.L. Interestingly, this constraint is mainly due to
a too strong enhancement of $r_{\gamma\gamma}$ for larger values of $x_u$.  On
the other hand, $\alpha$ is essentially unconstrained now, except that very
small mixing angles are disfavored. The latter follows from $y_u=0$, which leads to
a strong suppression of $r_{ZZ}$ and $r_{WW}$ when the mixing becomes too
small. 

The $y_d$ dependence shows an inverted behavior. Values above 1.6 increase the
total width and thus suppress both
$r_{ZZ}$ and $r_{\gamma\gamma}$, while very small values lead to a too strong
enhancement of both channels. Together this suggests some correlation in the
$x_u$--$y_d$ plane, as can be seen in the right plot of Fig.~\ref{fig:res3}. The
region that leads to good agreement with the data is relatively constrained. The
modifications of $r_{\gamma\gamma}$ and $r_{ZZ}$ cancel along certain diagonal
directions, but they start deviating from each other for larger values of
the couplings. In addition, $r_{\tau\tau}$ grows with $x_u$ and, to a lesser
extent, with $y_d$, such that very large values of both couplings are also
disfavored by this observable.

The correlations in the $x_d$--$y_d$ and $x_u$--$y_u$ planes are very similar to
the case of the single resonance model, so we do not show them separately.
Overall, we find that the double-resonance model imposes stronger constraints on
the variations of the Yukawa couplings, while the mixing angle is less
constrained in this scenario. In addition, the possibility to enhance
$r_{\gamma\gamma}$ without modifying the Yukawa couplings of the CP-even
component sets this model apart from the single-resonance case.

%
%

\subsection{Effective Theory Including New Dimension-five Operators} 

Now we turn to a discussion of the effects of higher-dimensional operators
which couple the CP-odd component $A$ to gauge bosons, see eq.~\eqref{dim5}, and
which are induced by loops of heavy new particles. The contribution of these
operators, with coefficients $c_G$, $c_W$, and $c_B$, comes in addition to
contributions from top and bottom quark loops, which can induce sizable
couplings between $A$ and $\gamma\gamma$ and $gg$ pairs through the couplings
$x_{\rm u,d}$. [The effect of top and bottom loops is generally negligible for
$ZZ$ and $WW$ pairs, which are dominated by the tree-level $HZZ$ and $HWW$
interactions.]

The expressions for the partial decay widths, including the contributions from 
both the modified top/bottom loops and $c_{G,W,B}$, are shown in 
eqs.~\eqref{eqn:gg5}--\eqref{eqn:zz5}. To avoid redundancy with the results of
the previous sections, we set in the following $x_u=x_d=0$ and $y_u = y_d=1$. In
the limit $\alpha=\pi/2$ we recover the model~\cite{Coleppa:2012eh}, a pure
CP-odd singlet scalar that only couples to the SM through dimension-five
operators. 

Letting the mixing angle and the coefficients of the dimension-five operators float, we find the minimum of the $\chi^2$ distribution at 
\begin{align}
	\alpha = 0.28\,, \quad c_G = 0\,, \quad c_B = 0.32 \,,\quad c_W = 11.5 \,.
\end{align}
With $\chi^2 = 8.86$, this provides a better fit than the SM, but slightly worse
than the scenarios considered in Sec.~\ref{1res} and~\ref{2res}. The dominant
effect is an enhancement of the di-photon rate to $r_{\gamma\gamma} \approx 1.7$ 
from the dimension-five operators. The small mixing angle, together with $c_G=0$,
suppresses all other rates by roughly 10\% compared to the SM. 
%
\begin{figure}
\begin{center}
\epsfig{width=0.45\textwidth, file=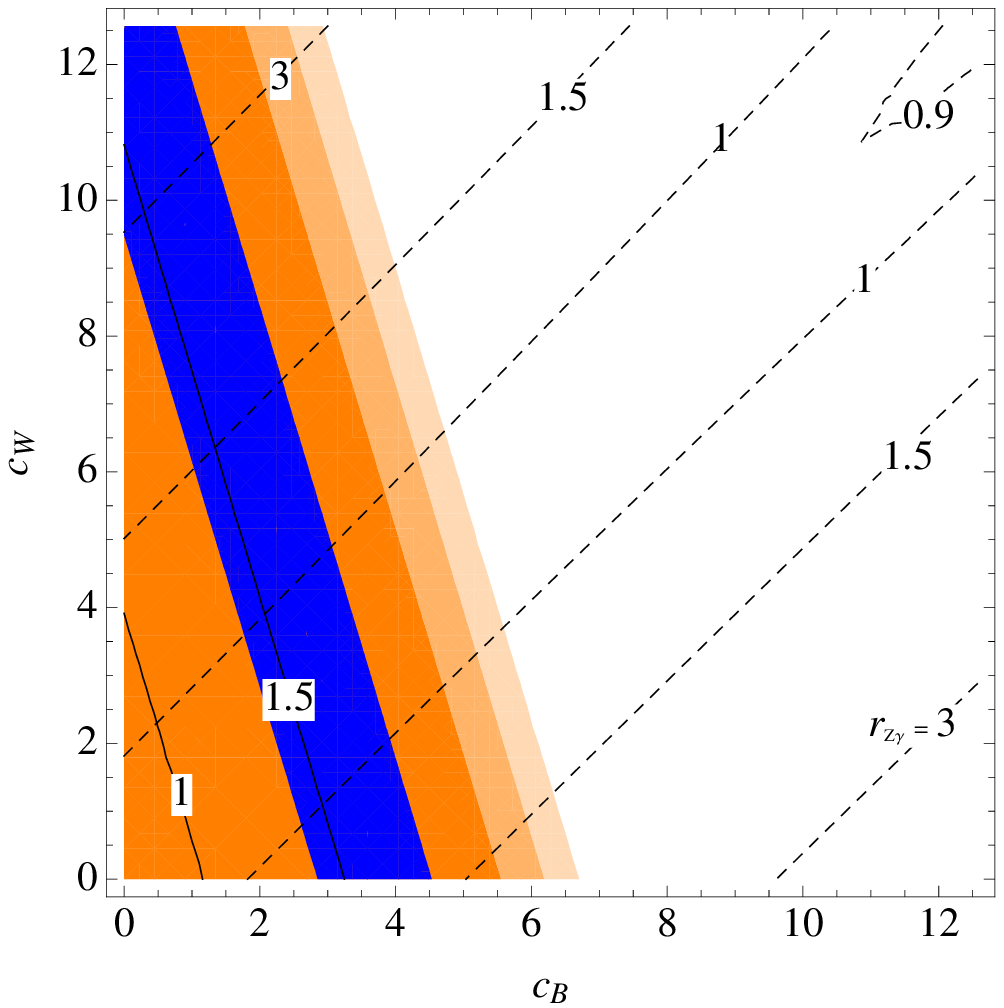}\hspace*{1cm}
\epsfig{width=0.45\textwidth, file=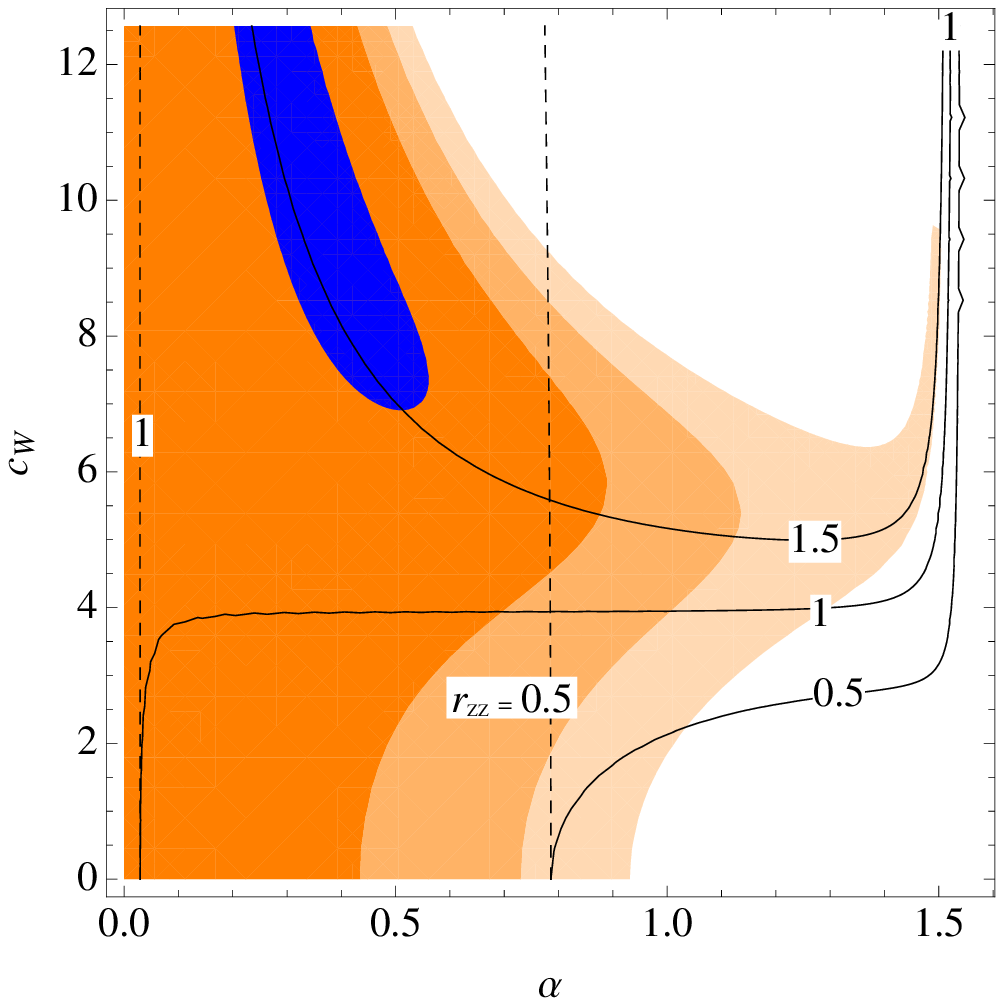}
\end{center}
\vspace{-1em}
\mycaption{Quality of the fit in the effective operator model, in the
$c_W$--$c_B$ (left) and  $c_W$--$\alpha$ (right) plane. 
The remaining free parameters are set to their best fit values, see text for 
details. As in the previous figures, the shaded
areas agree with the data at the level of $3\sigma$ (light), $2\sigma$ (medium),
$1\sigma$ (dark), and $\Delta \chi^2 < 1$ (blue/very dark). The solid lines are
contours of constant $r_{\gamma\gamma}$, while the dashed lines denote constant
$r_{Z\gamma}$ (left) and $r_{ZZ}$ (right).}
\label{fig:res4}
\end{figure}

%
Since only $r_{\gamma\gamma}$ is notably modified, the couplings $c_B$ and $c_W$
are not probed separately but only through the combination $c_B + \cot \theta^2 c_W$.
This is immediately obvious from Fig.~\ref{fig:res4}, where the iso-contours of
$r_{\gamma\gamma}$ in the $c_B$--$c_W$ plane are straight lines. To probe the
couplings individually one would have to measure the ratio $r_{Z\gamma}$, which
is proportional to $c_B$--$c_W$. A combined measurement of both rates would then
single out a circular region in the $c_B$--$c_W$ plane. However, the required 
precision in $r_{Z\gamma}$ can only be achieved with higher luminosity at the
14~TeV LHC~\cite{zgamma2}. 

In the absence of a result for $r_{Z\gamma}$, we can set $c_B=0$ without loss of
generality and analyze the constraints on the mixing angle in this model. In 
the right plot of Fig.~\ref{fig:res4} it can be seen that very large mixing
angles, up to $\alpha \approx 1$, are compatible with the data, but smaller
mixing angles are slightly preferred. The strongest constraints on the mixing
angle come from measurements of $r_{ZZ}$ and $r_{WW}$, since those rates 
decrease as $\cos^2 \alpha$ with increasing $\alpha$. 

So far we have kept $c_G = 0$, which is favored by the global fit. Obviously, for
very large mixing angles $\alpha \sim \pi/2$ this leads to disagreement with the
data since $r_{ZZ,WW} \to 0$ in this regime. To estimate the viability of the
pure CP-odd scenario, we can instead fix $\alpha = \pi/2$ and redo the fit.
Since $\Gamma(\phi \to b\bar{b})$ is zero here, the dominant decay of $\phi$ is
into  gluon and photon pairs and into $Z\gamma$, while the $WW$ and $ZZ$ decays are
suppressed by the three-body phase space. A realistic value for both
$r_{\gamma\gamma}$ and $r_{ZZ}$ then requires an approximate cancellation of the
effective photon coupling: $(c_\theta^2 c_B + s_\theta^2 c_W)^2 \ll c_B^2,\,
c_W^2$. 

It is still impossible, however, to achieve $r_{ZZ} \sim 1$ if the operator
coefficients are restricted to $|c_i|< 4\pi$, as suggested by perturbativity.
While the measured $r_{\gamma\gamma}$ is well reproduced, the model is excluded
at the 99.7\% C.L. due to the absence of a signal in all other channels. Note
that allowing for nonzero $x_u$, $x_d$, $i.\,e.$ allowing the CP-odd component to
couple to SM fermions, does not lead to an improved fit. The reason is that
while a nonzero $x_u$ can increase the production cross section, it also
increases $\Gamma(h \to gg)$ (and therefore the total width of $\phi$) so that the 
effects drop out in the ratios $r_{ZZ}$ and $r_{WW}$. We therefore arrive at the
very strong conclusion that a pure CP-odd resonance is excluded at the $3\sigma$
level in any perturbative extension of the SM. 

Finally, note that our setup also allows us to study the fermiophobic limit
$x_{u,d} = y_{u,d} = 0$. In the absence of mixing, this parameter point 
disagrees with the data at the $3\sigma$ level. Once mixing and nonzero
coefficients for the dimension five operators are allowed, we instead find a
good fit to the data for
\begin{align}
	\alpha = 0.84 \,,\quad c_G = 0.94 \,,\quad c_B = 0.47 \,,\quad c_W = 0.16\,,
\end{align}
with a $\chi^2$ similar to the SM fit. The Tevatron evidence for Higgs to $b\bar{b}$ decays is in tension with this parameter region. However stronger evidence of Higgs decays to SM fermions is required to probe a mixed fermiophobic scalar. 

%

\section{Future Projections}
\label{future}

It is worth estimating how much the constraints on CP violation in
Higgs mixing can be improved with future data from the LHC, using only the rate
information.  We expect that a full analysis of the 2012 8-TeV data set will
lead to a reduction of the uncertainties by roughly a factor two. 

In the long term, measurements in the 14 TeV run will not only improve the
sensitivity in the current search channels, but will further add rate
measurements in not yet observed channels. Specifically, we have considered the
following additional channels from Ref.~\cite{atlashigh}:
\begin{itemize}
	\item $h\to W W^*$ (VBF tag)
	\item $h \to \mu \mu$ (inclusive)
	\item $Vh$, $h\to \gamma\gamma$
	\item $t\bar{t} h$, ($h \to \gamma\gamma$ and $h\to \mu\mu$)
\end{itemize} 
In addition, the note specifies projected sensitivities for the $ZZ$,
$WW$~(inclusive), $\gamma\gamma$~(inclusive), $\gamma\gamma$~(VBF), and
$\tau\tau$~(VBF) channels, which are also included in our estimate. Projections are not given for the $b\bar{b}$ and
$Z\gamma$ channels. 

\begin{figure}[t]
\begin{center}
\epsfig{width=0.45\textwidth, file=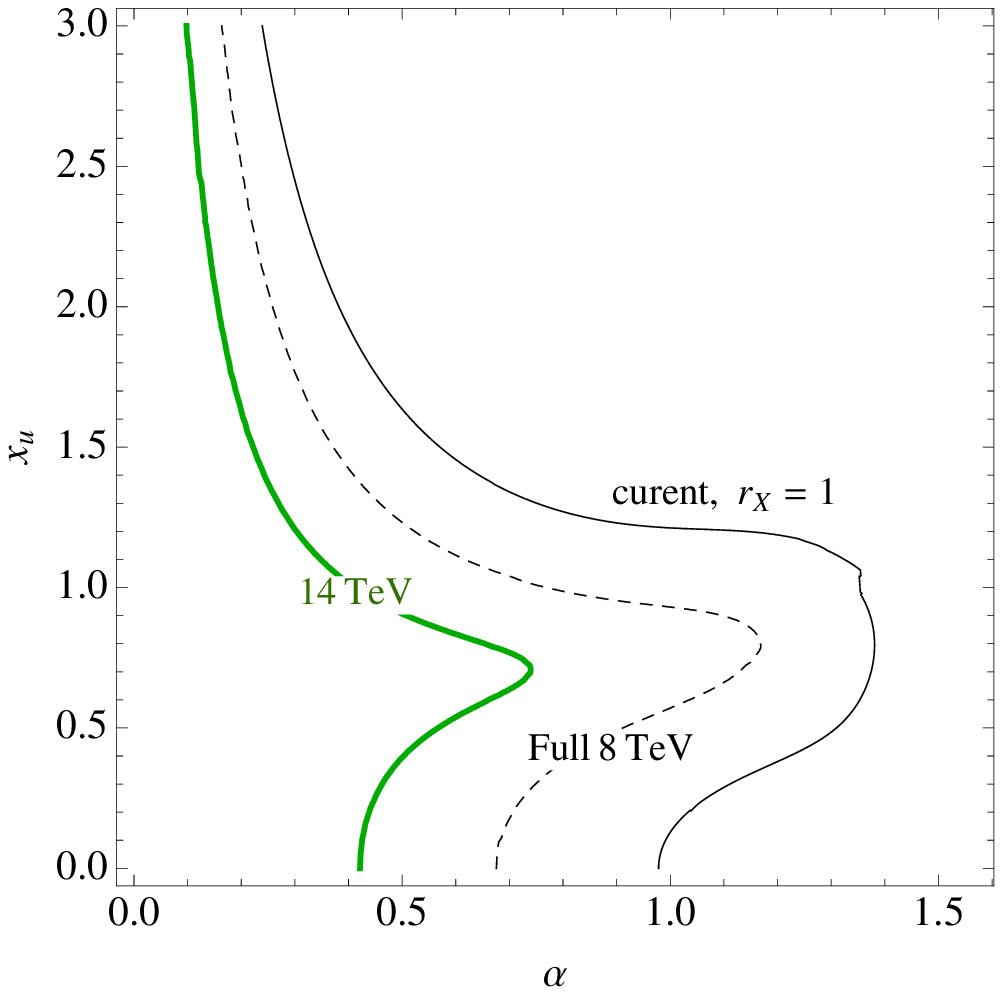}
\end{center}
\vspace{-1em}
\mycaption{Projected sensitivity on the CP mixing angle $\alpha$ and CP-odd top Yukawa
coupling $x_{\rm u}$ from upcoming LHC data on Higgs rate measurements. 
The plot is similar to Fig.~\ref{fig:res1}, under the assumption that all rate 
measurements have a central value consistent with the SM. Shown are the 95\%
C.L.\ limits for current errors but SM-like central values (light solid),
quadrupled statistics per experiment at the end of the 8~TeV run (light dashed), and expected
errors for 300~fb$^{-1}$ at 14~TeV
including only the channels discussed in~\cite{atlashigh} and combining the two
experiments (green, thick).}
\label{fig:resX}
\end{figure}

Of course, it is impossible to predict if the central values will remain the same
or shift when more data is analyzed. For concreteness,  it is interesting
to estimate how well the mixing angle can be constrained under the assumption
that the rate measurements will converge towards the SM predictions. To
illustrate this, in Fig.~\ref{fig:resX} we show three curves in the $\alpha$--$x_u$
plane. As in Fig.~\ref{fig:res1}, we set $x_d=0$, $y_u=y_d=1$, and
$c_G=c_W=c_B=0$. However, instead
of using the currently measured signal strengths, we assume that $r_{X,{\rm
measured}}=1$ for all channels. The three curves then are the 95\% C.L.\ limits
given \emph{(i)} the current error bars (light solid), \emph{(ii)} assuming a factor two improvement
on the error bars (light dashed), and \emph{(iii)} expected uncertainties based on the $300~{\rm fb}^{-1}$ 14~TeV ATLAS
projections and including an additional factor of two in the statistics assuming
that CMS reaches similar sensitivity  (green, thick). Note that for the last curve, we only include the channels for which projections are given in~\cite{atlashigh}. 

As evident from the plot, the mixing angle $\alpha$ in this scenario could be
constrained to $\alpha \lesim 1.1$ at the end of the 8~TeV run, and to $\alpha
\lesim 0.7$ with data taken at full energy and luminosity. The bounds  will thuse
become significantly stronger, although they are weakened by the dependence on
other parameters such as $x_{\rm u}$.

With larger available data sets it becomes feasible to constrain CP properties
from angular distributions, and the projected sensitivity at the end of the
8~TeV run is similar to Fig.~\ref{fig:resX} \cite{zzCP8}. There are currently no
estimates for the angular analysis of decay products of a 125-GeV boson at
14~TeV, but it is likely that it will lead to superior limits compared to
Fig.~\ref{fig:resX}, since this method is not affected by the \emph{a priori}
unknown couplings $x_{\rm u,d}$.


\section{Conclusions}
\label{concl}

The CP properties of the newly discovered boson with mass in the range
125--126~GeV can already be constrained with existing data on production rates
and branching ratios. The main reason is that a CP-odd pseudoscalar generally
has suppressed couplings to $W$ and $Z$ bosons since such couplings are generated only by 
higher-dimensional operators. In this paper, the 125-GeV resonance $\phi$ has been
assumed to be a general mixture between a CP-even and CP-odd scalar field, and
bounds on the mixing angle have been derived in a variety of different 
scenarios:
\begin{itemize}
\item[\emph{(i)}] fermion Yukawa couplings fixed to their Standard Model values;
\item[\emph{(ii)}] the overall scale of up-type and down-type Yukawa couplings to the
CP-even and CP-odd components may float freely and independently;
\item[\emph{(iii)}] the signal peak near 125~GeV is
comprised of two particles, $\phi$ and $\phi'$, which are the two mass eigenstates of the mixed
CP-even and CP-odd scalar fields;
\item[\emph{(iv)}] addition of higher-dimensional operators that couple the CP-odd
component to gauge bosons;
\item[\emph{(v)}] a special case of \emph{(iv)} with vanishing Yukawa couplings
(fermiophobic scalar).
\end{itemize}
Using the most recent Higgs search results released by ATLAS and CMS in
July 2012 \cite{atlash,cmsh}, it turns out that the possibility that $\phi$ 
is a pure pseudoscalar is already excluded at more than three
standard deviations, assuming that the new-physics sector is weakly coupled.
Nevertheless, large values of the CP-mixing angles, $\alpha \sim 1.0\dots 1.3$, are still
allowed at the 95\% confidence level, although smaller values are in better
agreement with the data. Interestingly, a non-zero mixing angle,
$\alpha \sim 0.15 \dots 0.4$, together with modified Yukawa couplings,
produces a slightly better fit than the Standard Model, although the
difference is not significant. If one allows the possibility that two mixed
scalars contribute to the observed resonance peak (scenario \emph{(iii)}), no
conclusive constraint on the mixing angle can be derived from current data.

Besides the mixing angle, meaningful limits on the Yukawa couplings and
coupling strengths of higher-dimensional operators of $\phi$ have been obtained
in all scenarios. The bounds on these parameters and on $\alpha$ are expected to
improve significantly with increased statistics, which will lead to more precise
measurements of the production rates and branching ratios, as well as open up
the possibility to constrain the CP properties by studying angular distributions
of the decay products.


\section*{Acknowledgements}

This project was
supported in part by the National Science
Foundation under grant PHY-1212635,  and by the U.S.\ Department of Energy,
Division of High Energy Physics, under contracts DE-AC02-06CH11357 and 
DE-FG02-12ER41811.


\section*{Appendix}

The Lagrangian~\eqref{dim5} leads to the following Feynman rules for the 
coupling of the CP-odd scalar $A$ to SM gauge bosons:
\begin{align}
        &A\gamma \gamma && \frac{c_\theta^2 c_B + s_\theta^2 c_W}{(4 \pi)^2 v} \epsilon^{\mu\nu\rho\sigma} (p_2 - p_3)_\rho (p_1)_\sigma \,,\\
        &AZZ &&  \frac{s_\theta^2 c_B + c_\theta^2 c_W}{(4 \pi)^2 v} \epsilon^{\mu\nu\rho\sigma} (p_2 - p_3)_\rho (p_1)_\sigma \,,\\
        &A\gamma Z && \frac{s_\theta c_\theta(c_W-c_B)}{(4 \pi)^2 v} \epsilon^{\mu\nu\rho\sigma} (p_2 - p_3)_\rho (p_1)_\sigma \,,\\
        &AWW && \frac{c_W}{(4 \pi)^2 v} \epsilon^{\mu\nu\rho\sigma} (p_2 - p_3)_\rho (p_1)_\sigma \,,\\
        &AGG && \frac{c_G}{(4 \pi)^2 v} \epsilon^{\mu\nu\rho\sigma} (p_2 - p_3)_\rho (p_1)_\sigma \,,
\end{align}
where and $p_{1,2,3}$ are the momenta of the first, second and third
particle flowing into the vertex. 


\end{document}